\documentclass[sigconf]{acmart}

\usepackage{booktabs, multirow} 
\usepackage{epsfig, graphicx, caption, subfigure}

\setcopyright{rightsretained}

\begin{document}
\title{Optimal Reserve Price for Online Ads Trading Based on Inventory Identification}


\author{Zhihui Xie}
\affiliation{%
	\institution{Alibaba Group}
	\city{Beijing}
	\postcode{100102}
	\country{China}
}
\email{zonglin.xzh@alibaba-inc.com}

\author{Kuang-Chih Lee}
\affiliation{%
	\institution{Yahoo! Inc.}
	\city{Sunnyvale}
	\state{CA}
	\postcode{94089}
	\country{U.S.A}
}
\email{kclee@yahoo-inc.com}

\author{Liang Wang}
\affiliation{%
	\institution{Alibaba Group}
	\city{Beijing}
	\postcode{100102}
	\country{China}
}
\email{liangbo.wl@alibaba-inc.com}

\copyrightyear{2017}
\acmYear{2017}
\setcopyright{acmcopyright} \acmConference{ADKDD'17}{August 14, 2017}{Halifax, NS, Canada}\acmPrice{15.00}\acmDOI{10.1145/3124749.3124760} \acmISBN{978­1­4503­5194­2/17/08}
 
\renewcommand{\shortauthors}{Z. Xie et al.}

\begin{abstract}

The online ads trading platform plays a crucial role in connecting publishers and advertisers and generates tremendous value in facilitating the convenience of our lives. It has been evolving into a more and more complicated structure. In this paper, we consider the problem of maximizing the revenue for the seller side via utilizing proper \emph{reserve price} for the auctions in a dynamical way. 

Predicting the optimal reserve price for each auction in the repeated auction marketplaces is a non-trivial problem. However, we were able to come up with an efficient method of improving the seller revenue by mainly focusing on adjusting the reserve price for those high-value inventories. Previously, no dedicated work has been performed from this perspective. Inspired by Paul and Michael \cite{Viola01robustreal-time}, our model first identifies the value of the inventory by predicting the top bid price bucket using a cascade of classifiers. The cascade is essential in significantly reducing the false positive rate of a single classifier.  Based on the output of the first step, we build another cluster of classifiers to predict the price separations between the top two bids. We showed that although the high-value auctions are only a small portion of all the traffic, successfully identifying them and setting correct reserve price would result in a significant revenue lift. Moreover, our optimization is compatible with all other reserve price models in the system and does not impact their performance. In other words, when combined with other models, the enhancement on exchange revenue will be aggregated. Simulations on randomly sampled Yahoo ads exchange (YAXR) data showed stable and expected lift after applying our model.

\end{abstract}


\maketitle

\pagestyle{plain}
\setcounter{page}{1}
\pagenumbering{arabic}

\section{Introduction}

Online advertising have been increasingly important for the market participants. In a report by Mckinsey \cite{McKinsey}, the digital ads marketplace has seen a strong growth and digital ads sale will reach \$231 billion by 2019. Before the emerging of ad exchange, there are many intermediaries in the value chain between publishers and advertisers each taking a slice of ad market share. Both the advertisers and publishers spend significant time working on where and how to buy and sell so that impressions go to the right audience. The ad exchange solves these problems and simplifies the process of serving users, advertisers and publishers in a much more efficient way. 

Yahoo's BRXD (Yahoo display ads exchange) is one of the major trading platforms that serves numerous advertisers and publishers. It enables the potential buyer to reach hundreds of millions of users across the world at a single platform, from which a pool of specialized targeting audiences is chosen according to the buyer's individualized campaign goals. The centralized trading structure enables small buyers who have limited budget to reach a large variety of users that would not be possible without a platform. In general, the advertising inventory is bought and sold on a per-impression basis by running a real-time auction. The system has evolved into a stage such that it is programmed to be able to run millions of auctions, serve the corresponding ads, record the payment, fight with fraud, ... all such kinds of procedures within one single second. In this paper, we are focused on a method to determine how bidders are charged at impression level. 

When calculating the transaction prices, the industry generally uses the so called popular standard \emph{second price auction} design with a reserve price:  a valid top bidder (the winner) will pay the second bidder's price or the reserve price, whichever is higher. The current BRXD has a very low global reserve price (a few cents) for all auctions, and are not supportive for dynamic scenarios. One challenge is that it is difficult to model bidder behavior and its evolution as auctions are running on various types of inventories. On one hand, over-predicting is quite dangerous since it would result in blocking of all good bids and publisher suffers from the opportunity cost. On the other hand, extremely conservative reserve price is useless since second price is already served as good price support. As a platform, Yahoo has substantial chunk of inventories owned by itself, thus a seller revenue improvement also benefits Yahoo's revenue performance directly.

\subsection{Second price auction}
The online marketplace of ads trading is similar to the traditional item auctions in terms of winning rule. There are extensive studies in the auction designs in the context of traditional economic regime. Driven by the goal to maximize revenue,  one may be wondering why the publisher can resist the temptation to charge the winner the top price instead of second price. 

One of the major findings of auction theory is the celebrated Revenue equivalence theorem. Early equivalence results focused on a comparison of revenue in the most common auctions. The first such proof, for the case of two buyers and uniformly distributed values was by Vickrey \cite{Vickrey61} in 1961. Later Riley \& Samuelson \cite{RileySam81} proved a much more general result. (Quite independently and soon after, this was also derived by Myerson \cite{RM81} in 1981). The revenue equivalence theorem states that any allocation mechanism or auction that satisfies the \emph{main assumptions} of the benchmark model will lead to the same expected revenue for the seller. The main assumptions includes bidders are risk-neutral, their valuation for the inventory is independently and identically distributed (i.i.d) and the payments must depend on bids. 

It is a well-known result that in first price auction, where the player with the highest bid simply pays whatever it bids, a Nash equilibrium is that each player bids such that everybody bids the expected value of second highest bid. This result is also based on the \emph{main assumptions}. In the real-time ads bidding environment, second price auction awards an impression to the ad with the highest bid, but records the revenue of the impression as the second highest bid. It is optimal for the bidders to bid at its own valuation of the impression under the same assumptions. For both auction mechanisms, the impressions are allocated to the same bidders with the same expected revenue, this is the famous revenue equivalence theorem in auction theory. From the bidders point of view, the second price auction is definitely more attractive than first price auction at first glance. However, the above main assumptions are not always true in realistic ads trading platform (actually it is almost never true for Yahoo's exchange). For instance, some bidders shared very similar bidding formula for certain inventories, it is hard to verify whether their bids are independent. Moreover, there are many types of bidders with different priorities, some put deliver rate over CPM (cost per thousand impression) goal and are willing to overbid; whereas some bidders are short of budget and prefer shading their bids. No one can guarantee that the new equilibrium will ensure a dominant strategy of truth bidding, see \cite{SantiagoB2013, BBW13}.

The idea of reserve price (also called \emph{floor price}) is to set up a lowest price for each auction. Instead of paying the second bid price, the winner will pay the higher of second bid and reserve price, as if the seller place a bid at the reserve price. This mechanism prevailed the online bidding market since it essentially provides a protection of transaction prices for the publishers. Setting a good reserve price for the inventory is an interesting question. If it is too high (greater than first bid), the inventory will be left unsold and incurs opportunity cost. However, if the reserve price is too low (lower than the second bid), it fails to protect the publisher from auctions with low second bid and high first bid prices. Ideally the reserve price should be between the top two bids:
\begin{equation*}
	\text{Second bid} < \text{Reserve price} \leq \text{Top bid}.
\end{equation*}

Besides the static reserve price, publishers can also set a variable price with a minimum and maximum, which enables an ad to win at lower prices for less valuable inventory and higher prices for higher value inventory.

\subsection{Past work and our method}  \label{subsec: pastwork}

Under the second price auction (also referred as \emph{Vickrey auction}) scheme, reserve price is critical to maximize revenue (we consider seller side revenue throughout this paper), this is proved by Myerson in 1981 in his classic Nobel prize winning work \cite{RM81}. Actually Myerson showed that if the valuations are drawn independently and identically from a distribution satisfying a regularity assumption, the optimal auction takes the form of a second price auction with a reserve price. Furthermore, if the bidders are risk neutral, Riley and Samuelson \cite{RileySam81} confirmed that the second price auction yields the same expected revenue as the first price auction. 

We noticed some recent works \cite{ARS13, MM14, CGM15} on the evolution of bidder's strategy under repeated auctions. People have different approaches to formulate and solve the optimization related problems. One of the pioneer work \cite{Mohri2014} formulate the optimal revenue problem as a learning problem and present a full theoretical analysis on it. \cite{DRY10} aims to achieve almost optimal expected revenue for arbitrary bidder valuation distribution.  The key idea in which is to associate each bidder with another that has the same attribute, with the second bidder's valuation acting as a random reserve price for the first. \cite{LPV16} studied lazy and eager versions of reserve prices, the idea in which is similar to another version of the soft reserve price. The optimization problem for the buyer side has been well studied, \cite{RepeatedMAB16} investigates the repeated version of Shubik's dollar auctions and discussed effect of different player types.

It seems at first sight that having a reserve price will definitely help improving the publisher yield in second price auctions. This is not always true, for instance, the impact of soft reserve prices in advertisers is studied in  \cite{YuanWangZhao13}, it ``puts advertisers in an unfavorable position and could damage the eco-system". In the same paper, the authors also evaluate the performance of bidding strategy. In \cite{BBW13}, the authors find that proper adjustment of the reserve price is key in achieving profitable for the publisher to try selling all impressions in the exchange before utilizing the alternative channel. Our efforts in finding the optimal reserve price in order to optimize the seller revenue is in this background, we come up with an efficient model that are completely different from those in previous literatures that works great for our platform.

Since the top two bids vary from auction to auction, a constant reserve price across all inventory types will not be granule enough. An easy way to add fluctuations is to perform manual adjustments to the reserve price based on certain features. When the top two bid prices exhibit high correlation with time, the manual way would somewhat improve the publisher revenue. Nonetheless, it is in general a low efficient way to have manpower involved in auctions that usually complete in several mini-second. A more efficient way is to set the reserve price for each auction automatically. 

Before we come up with this data-driven method, we have already implemented a parametric model that works pretty well in our video platform \cite{Miguel} but not as expected in the display marketplace. In \cite{Miguel}, the authors use a game theory based parametric method that is very effective in computing the reserve price for sparse and high ranged auctions, such as the video marketplaces, where the bid price can be several dozens of dollars and number of bidders for each auction is small. The parametric model \cite{Miguel} assumes the bids follow some certain distributions, and historical bids data is used to fit the parameters. Under the benchmark assumptions a closed form solution can be derived by solving a differential equation. As we have discussed above, the benchmark assumptions is in doubt and it is hard to evaluate how much impact it could bring to the accuracy of the model. Thus the \emph{true} distribution of the bidder's valuation is very difficult to obtain. For BRXD, bids are much more dense and pricing range is narrower than video marketplace. One problem we found is that the parametric model is hard to achieve statistical significant revenue lift anymore.

Since the display marketplace has a higher capacity and are more competitive than video marketplace, there are all kinds of players that contribute a lot more noises. We started by not assuming any known type of distributions for the bids, but just assume that the top two bids are \emph{relatively} stable for certain type of inventories. This is a fundamental difference from the parametric approach and is in a more data driven spirit. Noticed that the optimal reserve price only depends on the top two bids, we will only focused on study the top bid and the gap between the top two bids, eliminating the requirement that all bids are independently identically distributed. 

To further reduce the impact of noise, we only recommend an effective reserve price for those inventories that are predicted to be \emph{high value} by our model, and leave the other ones' effective reserve price (see Section \ref{subsec: multiple reserve}) unchanged. We say that an inventory is of \emph{high value} if its potential top bid is greater than some threshold. The motivation for the choice of operating only on high value inventory is that we can achieve very good AUC with the features in our hand, and the implementation is relatively easy. Since the revenue margin narrows down when the top two bids are close, we also build a cluster of classifiers to estimates the price separation between the top two bids. For instance, if the price separation is small, the ROI (risk over return) for implementing an aggressive reserve price will be very low. We use the predicted separation as an indicator to determine whether an aggressive reserve price should be enforced or not.

\section{Features and Inventory Classification Buckets}     \label{sec: features}

\subsection{Multiple Reserve Prices}  \label{subsec: multiple reserve}

In online auction design, the reserve price is set before the bids come, any bids below the reserve will be blocked before enter into the ranking algorithm. We call this type of reserve price \emph{hard}. For a comprehensive ads exchange platform, there can be multiple types of hard reserve price, such as \emph{systemwide reserve prices}, \emph{uniform reserve price}, \emph{deal reserve price}, etc. 

The uniform reserve price is the minimum clearing price for certain ad dimensions that mainly depending on ad size or site. Deal reserve price is associated with deal seat, these are pre-negotiated before campaign set up. Usually the lowest one is the system wide reserve price, roughly a few cents. In the exchange, we put requirements in response rate and other KPIs for DSP. In order to secure their spots in the system, some DSPs are submitting extremely low bid for a large amount of auctions that they are not interested at all. The existence of system wide reserve can help reduce the number of bids considered for certain auctions and latency performance. The final actual effective reserve price is in general the maximal among all above static reserve prices that are quite conservative and are updated in a relatively long period based on platform statistics. 

Some platforms also use \emph{soft reserve price}, these type of reserve does not block any bids, but just provides a support for the transaction price. It will not incur any opportunity cost if the winner bids below the soft reserve. Actually, when underbidding (top bid is lower than reserve) happens, the auction continues as if there is no reserve price. It is also common to use a combination of soft and hard reserve prices, if the highest bid is above the hard reserve price but below the soft reserve price the auction is run on first price and the winners pay the exact amount they bid. One of the advantages of soft reserve prices is that it won't bring opportunity cost when underbidding, however it it hard to predict the bidder behavior. Long termly speaking, bidders may find they can win auctions at low price since underbidding does not result in losing auctions. One possible effect what goes against seller's interest is that bidders keep shading their bids and push the average clearing price to lower levels.  
 
In this paper, we will discuss how to find the optimal hard reserve prices at impression level. We start by collecting data from the exchange and user profile. 

\begin{table}
\caption{A list of inventory characteristics that determines the intrinsic value of a inventory.}
\label{table: features}
 \begin{tabular}{ |l|l|}  
\hline
\multirow{2}{*}{Publisher info} & ad section, site - top level domain (TLD), layout, \\
 & ad size, supply, ssp host, ad position, etc \\ \hline
\multirow{5}{*}{User info} & age, gender, device type, geo info, app info, \\
& browser, colo (SSP host location), page view, \\
& previous clearing price, \# of visit, conversion, \\ 
& impressions, clicks, previous winning statistics, \\
& search query(click), etc \\ \hline
\multirow{2}{*}{Buyer info} & group of buyer (DSP group identification), \\
& winning demand seat (lagged per user), etc \\ \hline
Other & date, hour of day, day of week (dow), etc \\
\hline
\end{tabular}
\end{table}

\subsection{Features}  \label{subsec: features}

We believe the value of an inventory depends both on its intrinsic characteristics and the buyers. From the raw data we filtered out some information that are ranked low for the price differentiation and left with Table \ref{table: features}.

Some of the characteristics in Table \ref{table: features} have multiple features, such as in the previous win statistics for a user. The user stat includes previous maximum, average, total count of previous prices, also the running average can be over a group of numbers.  One advantage of our model is that it does not rely on information about the losing bids. This gives more flexibility of the model. If we only have DSP data, we can still run the model. 

We categorize some features such as section id, site id and ad positions. Some feature are generated by grouping or binding and put into buckets. To name a few, we put thousands of buyer seats into 10 group of DSP buyers as AdNetwork, AgencyTradingDesk, DSP, DSPPowered, PersonalizedRetargeter, Adx, Gemini, Sidekick, YamPlus and notag. The \emph{age} is grouped according to the following buckets: 0-17, 18-20, 21-24, 25-34, 35-44, 45-54, 55-64, 65+.

The \emph{dow} and \emph{date} may help us reflect some periodic trends (weekly, quarterly, etc).  We also cleaned the outlier bids that are higher than \$41 (roughly 0.1\% percentile) before feed the model. 

\begin{figure}
    \hfill
    \subfigure[High-value identification]{\includegraphics[width=1.65in]{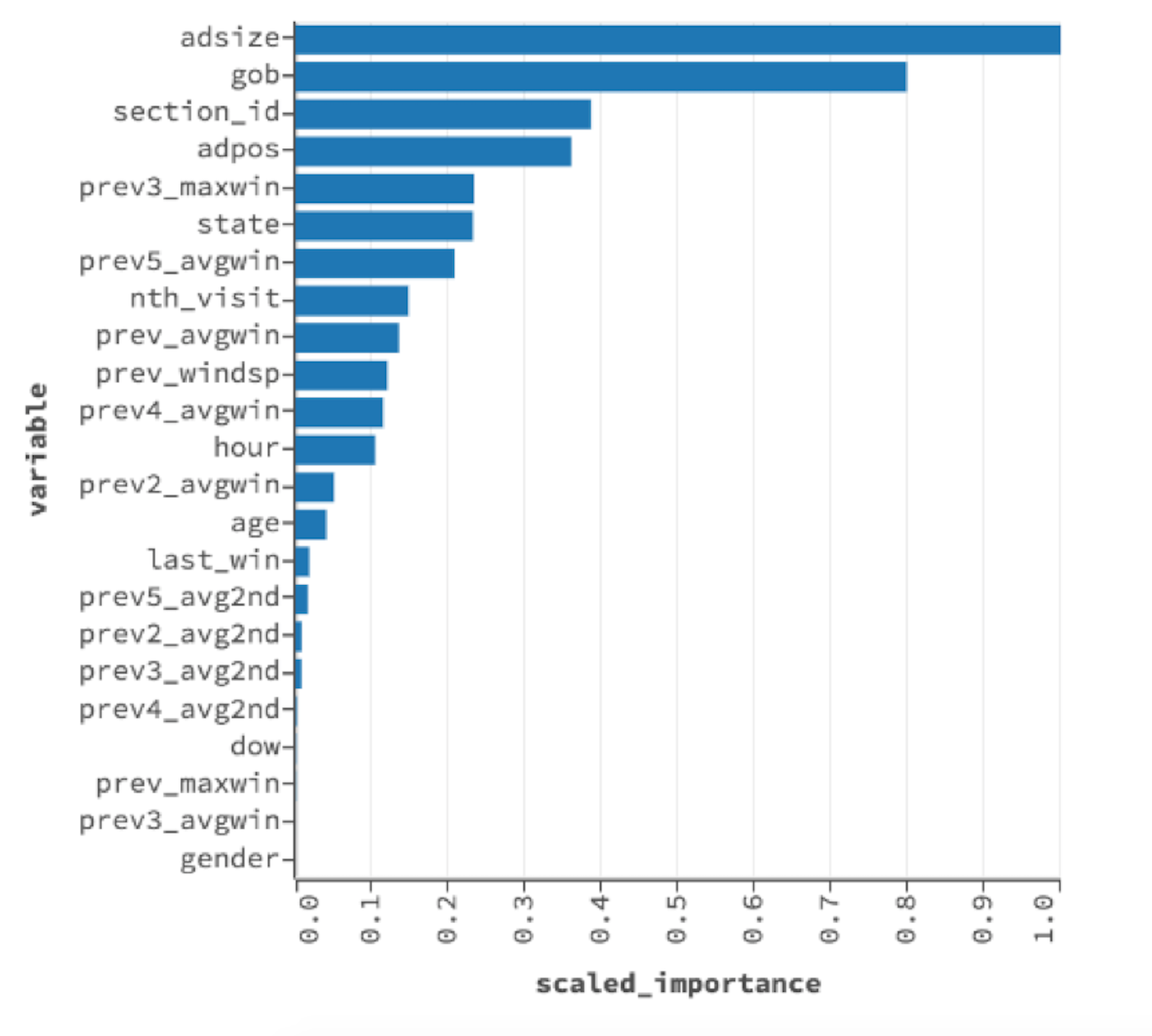}}
    \subfigure[Price separation detection]{\includegraphics[width=1.65in]{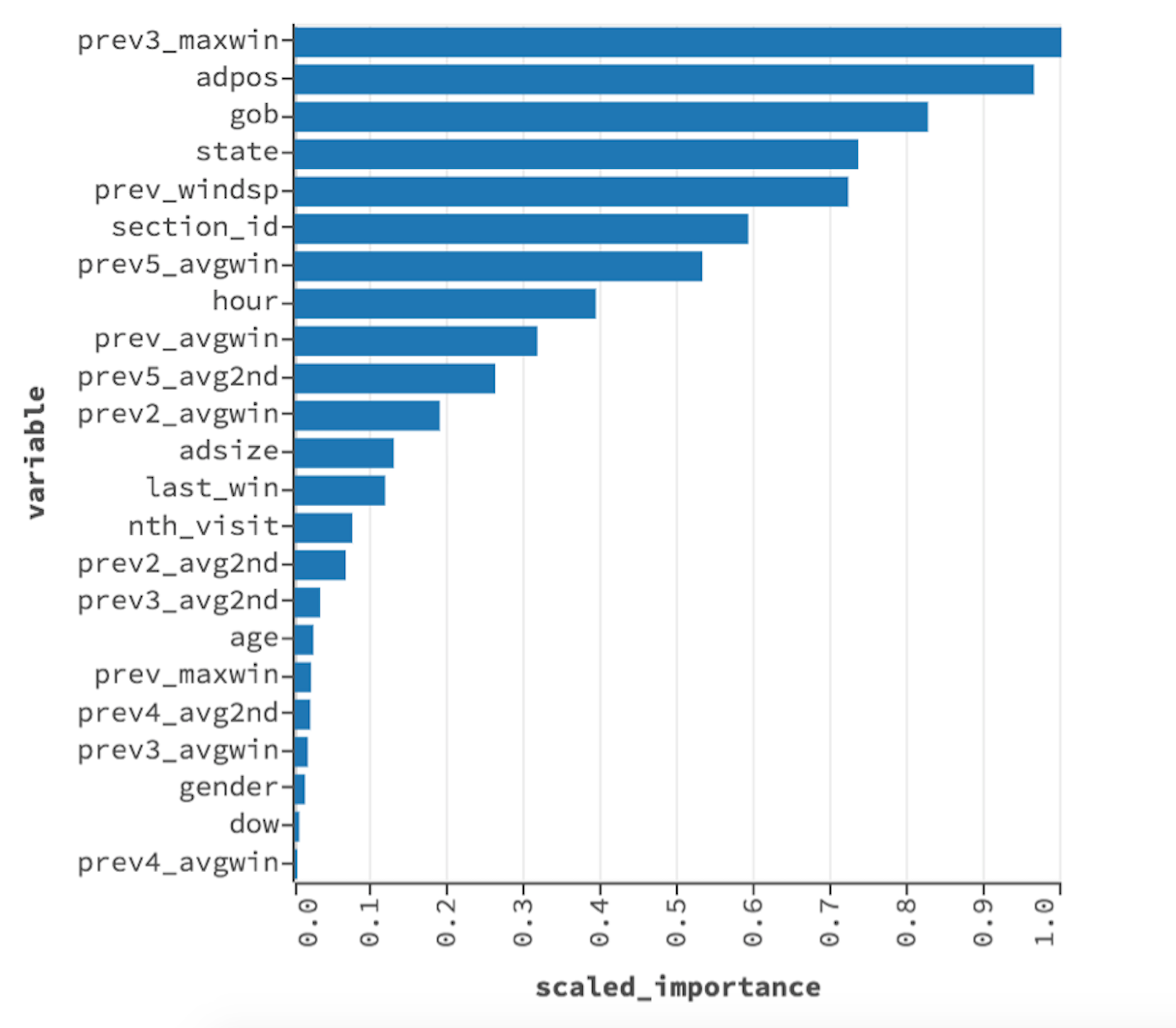}}
    \hfill
    \caption{Rankings of feature importance according to high-value and price separation classifiers.} \label{feature rank}
\end{figure}

A typical feature ranking is listed in Figure \ref{feature rank} for the two family of classifiers. From which we can see that the very top feature groups are highly correlated but with some internal difference. This is reasonable since the highest bid and the huge bids gap are both random variables over the same bidding pool, bidders share a lot of common information from the bid request which determines the intrinsic value of the inventory. The differentiation of bid prices are more relying on the bidder's interpretation and goals. 

\section{Model}  \label{sec: model}

\begin{figure}  
\includegraphics[height=1.75in, width=2.25in]{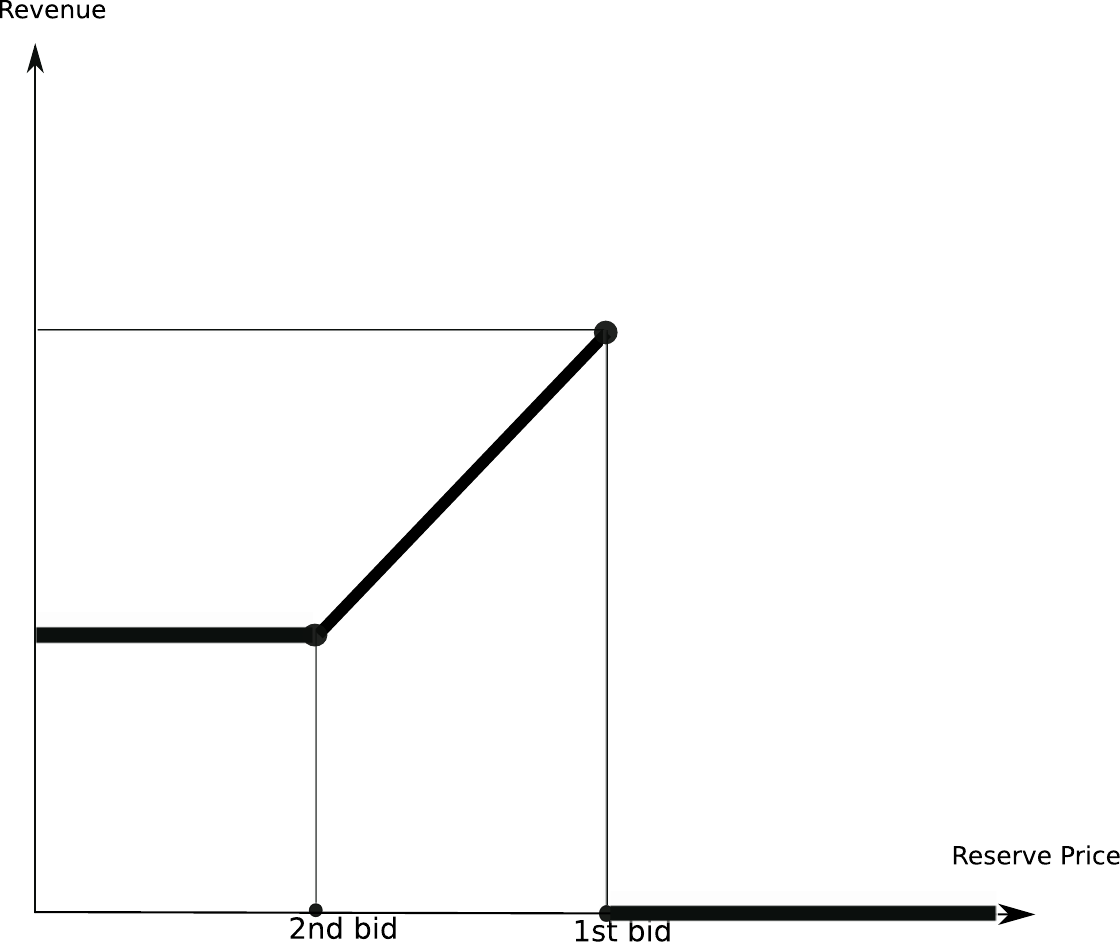}
\caption{With top two bids given, the seller revenue function at different reserve price: bold segments} \label{rvnu vs reserve}
\end{figure}

One challenge in calculating the dynamic optimal reserve price is the opportunity cost. Let us assume there is no resell, consider a simple auction with top bid $T=\$5$ (cpm, same for below) and second bid $S=\$2$ respectively. When reserve price $r$ is below \$2, the impression will be sold at \$2;  if $2\leq r \leq 5$ the winner will pay $r$; once $r>5$ there will be no sell and the seller suffers an inventory loss. We can easily see that the ideal $r$ should be between $T$ and $S$ and the higher the better. However, this linear dependency between the revenue and $r$ ends by a jump at $T$ (see the thick black curve in Figure \ref{rvnu vs reserve}). Start from a conservative $r$, in the pursue of maximal revenue, $r$ is pushed to be close to $T$. Since $T$ can be over predicted (false positive) while the true top bid $T$ is actually lower, an aggressive $r$ may result in revenue dropping to zero for the particular auction. So the practical decision process will involve balancing of prediction accuracy and opportunity cost, and tuning between risk and expected revenue. 
 
In our model, the calculation of $r$ is based on our prediction of discretized $T$ and their gap $T-S$. We essentially build several families of classifiers which are pre-trained using historical data. Each classifier is a cluster of regular machine learning models that are dedicated for fixed feature set and dimensions of the response column. Although there are multiple buckets after discretizing the winning price $T$, we are only interested in those high value buckets, thus the identification of high-value inventory is a binary decision. We also convert the gap $T-S$ to a binary value ($1$ for significant $0$ for not significant) by comparing it with a threshold. Further, we will not calculate new reserve price for inventories that does not have \emph{significant} gap between the top two prices.

\emph{Price separation prediction. } 
The first family of classifiers are \emph{separation classifiers}. We are aiming of identifying the price separations between the first bid price and second bid price. The model will only calculate the reserve price for auctions that are labeled positive by separation classifier. 

\emph{Winning bucket prediction. }
The second family of classifiers are \emph{prediction classifiers}. Our new effective reserve price is calculated based on the output of these classifiers. In the model, we discretize the winning price into different buckets, and the output will be the bucket ID that winning bid belongs to. Based on which we build the \emph{high value detection classifiers} which collects votes from this prediction classifiers. The final calculation of our reserve price depends on this bucket size.   

The two step model includes building of the single layer classifiers and the cascading algorithm. The cascading algorithm is inspired by Paul Viola and Michael Johns 's classic work \cite{Viola01robustreal-time} in image processing, whose method achieves both high efficiency and low false positive rate.

\subsection{Single Classifier Performance}

In the cascade procedures, we build strong classifiers in each stage. We pick a subset of features and train classifiers $h_j$ (weak learners) with the specific feature set, then we use boosting \cite{freund1997} (developed by Freund and Schapire in 1990s) to combine these weak learners. The feature set varies for different learners. The boosting process essentially selects a set of good learners which nevertheless have significant variety. It is old but works very effective in our case, the idea of the algorithm is to iteratively assign appropriate weights on each training examples (actually more weight on those training examples that are misclassified by the current weak hypothesis just produced and less weights on those training examples that are correctly classified so that the learner can focus on the hard to classify examples in the next iteration). Basic steps of this algorithm is as follows:

Let $T$ be the number of loops, at stage $t$, we assign weight $W_t(\cdot)$ to the training example $i$. The goal of the learning algorithm is to produce a hypothesis $h_t(\cdot)$ such that the error at current stage is minimized: $$\epsilon_t=\sum_{i=1}^n W_t(i) \mathbf{1}_{\{h_t (x_i)\neq y_i\}}.$$
Where $(x_i, y_i)$ are the training data, $\mathbf{1}_{\{\cdot\}}$ is the indicator function, it equals to $1$ if the argument in bracket is true, $0$ otherwise. $\epsilon_t$ is the probability of error of classifier $h_t(\cdot)$, with respect to the distribution of weights $W_t(\cdot)$. Since the weights are sum to be $1$, thus this error is always in $[0,1]$.

Upon finishing of all the iterations (see Table \ref{boosting code}), we combine the classification rules $h_t$,  $t=1,2,..., T$ and obtain a strong classifier $$H(x) = sign \Big(\sum_{i=1}^T \alpha_t h_t(x) \Big), $$
$sign(v)$ is the sign function, it returns $1$ if $v > 0$, $-1$ otherwise. The coefficients $\alpha_t$ are generated according to the errors as $\alpha_t = \frac{1}{2}\ln(\frac{1-\epsilon_t}{\epsilon_t}).$ For any given $x$, $H(x)$ will return a binary outcome depends on the sign of the weighted sum. 

\begin{footnotesize}
\begin{table}
\caption{Pseudo code for performing boosting procedure.}
\label{boosting code}
 \begin{tabular}{ll}  
\hline
1& Let $T$ classifiers be $h_1, h_2, \cdots, h_T$ \\
2& Training data $(x_1, y_1), \cdots, (x_n, y_n)$, $y_i\in \{+1, -1\}$ \\
3& Initialize $t=1$, $W_t(i)=\frac{1}{n}, \forall i$ \\
4& For $t = 1, 2, \cdots, T$:  \\
5& $\quad$ use distribution $W_t$ to obtain classifier $h_t$ \\
6& $\quad$ calculate the error $\epsilon_t$ and coefficient $\alpha_t$ \\
7& $\quad$ update the weights for next iteration as $W_{t+1}(i) = W_t(i)e^{-\alpha_t y_i h_t(x_i)}$ \\
8& $\quad$ normalize $W_{t+1}(i)$ so that it is a probability distribution, i.e., sum to be $1$ \\
9& Calculate the final output $H(x)$ \\
\hline
\end{tabular}
\end{table}
\end{footnotesize} 

Another reason we choose boosting is that the training error approaches $0$ exponentially in the number of rounds $T$.

\begin{figure}
    \hfill
    \subfigure[Training AUC = 0.909]{\includegraphics[width=1.65in]{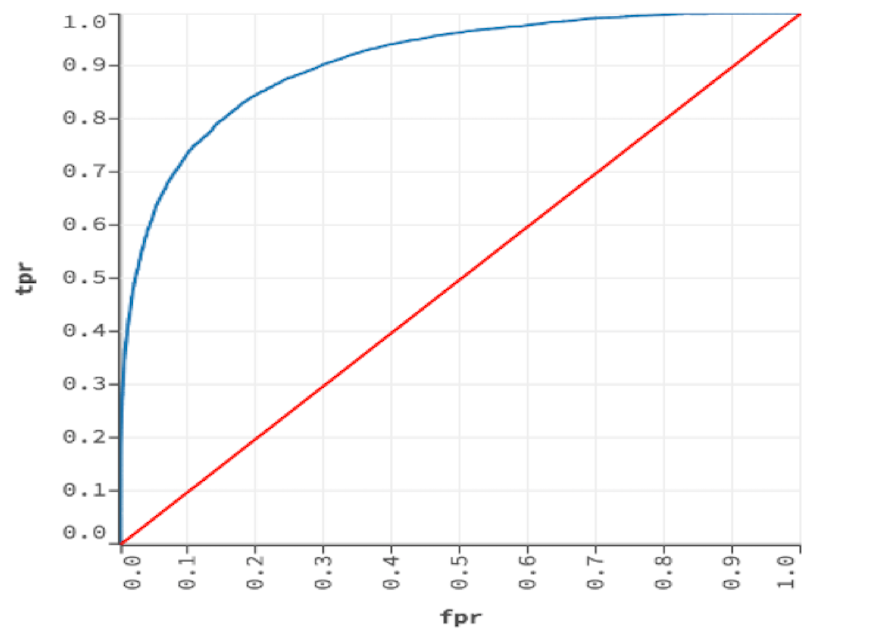}}
    \subfigure[Test AUC = 0.886]{\includegraphics[width=1.65in]{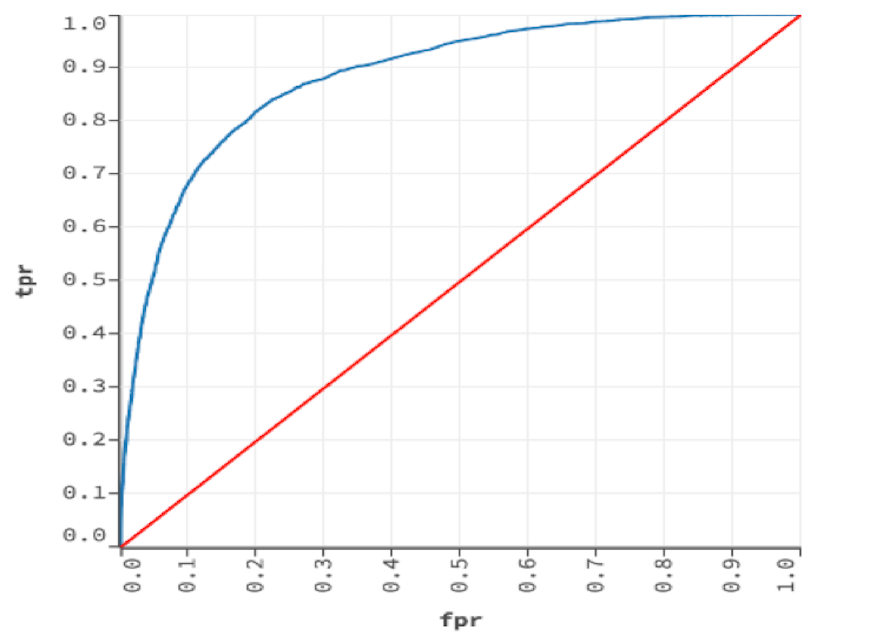}}
    \hfill
    \caption{ROC curve for high value identification classifiers.} \label{high value auc}
\end{figure}

\begin{figure}
    \hfill
    \subfigure[Training AUC = 0.785]{\includegraphics[width=1.65in]{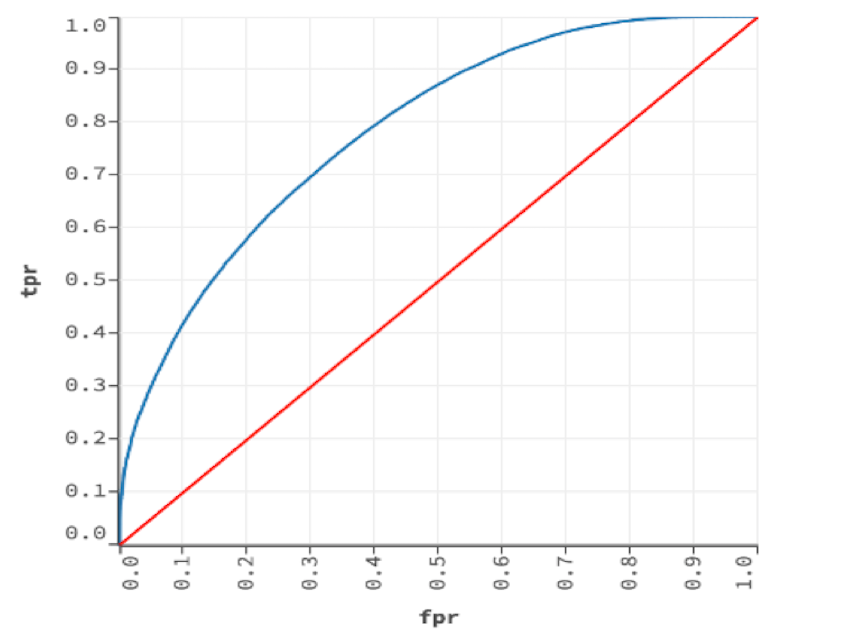}}
    \subfigure[Test AUC = 0.753]{\includegraphics[width=1.65in]{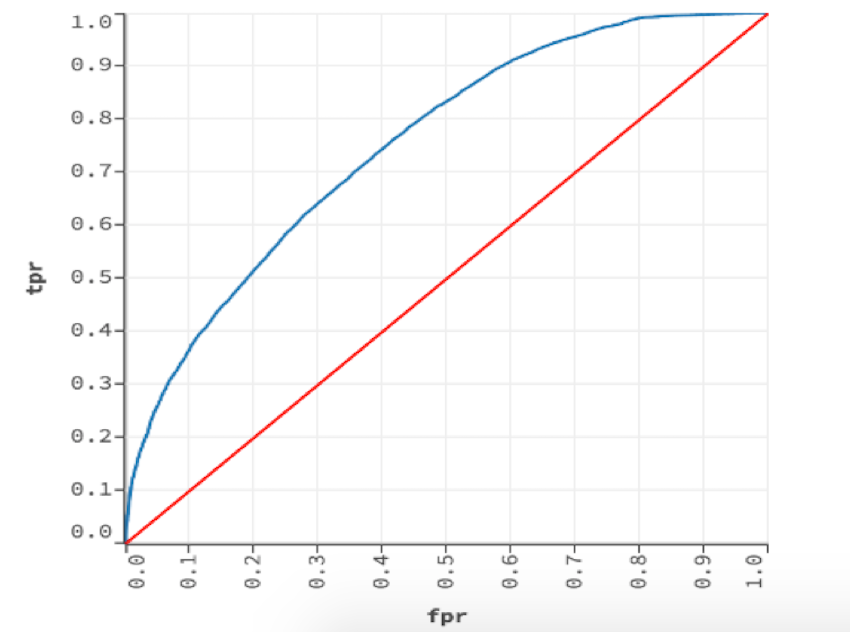}}
    \hfill
    \caption{ROC curve for price separation detection classifiers.} \label{price separation auc}
\end{figure}

The performance of single classifiers for either high value identification and top prices separation are pretty good (see Figure \ref{high value auc} and \ref{price separation auc} for the worst classifiers), especially for the high value identification classifiers, we have test AUC above $0.88$. One may argue that AUC performance does not directly translate to revenue because we do not know enough about the potential loss on the false positive samples. This is a serious concern and we did keep an eye on the projected revenue after finishing training and testing of our classifiers, the detailed revenue impact results can be found in Section \ref{sec: simu results}.

\subsection{Cascading Algorithm}   \label{subsec: cascade}

As of now we have a bunch of classifiers for both high value detection and price separation. The goal of cascading \cite{Quinlan1986} is to reduce the false positive rate of a prediction algorithm by combining a series of classifiers. We followed the basic idea in \cite{Viola01robustreal-time} with our own features and classification models. In order for a sample to be predicted as positive, all the classifiers in the cascading must predict it to be positive, otherwise it will be predicted as negative sample. In the training process, ``a positive result from current classifier triggers the evaluation of the next classifier which has also been adjusted to achieve high detection rate. A positive result from the next classifier triggers a third classifier and so on". A negative outcome from any point leads to a termination of the chain and labelled as negative immediately.  Subsequent classifiers are trained using the examples that pass through all the previous stages. And it is more and more difficult for the later classifiers to predict. The process is illustrated below, a series of classifiers are applied to each sample that passes all previous classifiers. Instead of putting simpler classifiers in the beginning, we choose smaller but important feature set for the classifiers at earlier stage so that they can eliminates a large number of negative samples in a relatively short time. Fewer and fewer samples survive after several stages of processing, and are handed over to more complicated classifiers. See \ref{cascade code} for pseudocode.

For a trained cascade of $N$ classifiers, both the detection rate ($D = \prod_{i=1}^N d_i$) and false positive ($F=\prod_{i=1}^N f_i$) rate declined almost exponentially. Our main goal is to reduce the false positive rate (corresponding to the market share) to certain level while maintain as much detection rate (corresponding to revenue) as possible. The single classifiers must maintain a relative high detection rate so that the goal is achieved.
For a $10$ stage cascade with each classifier has the same detection rate 0.95 and false positive rate 0.52, the overall detection rate will be $0.95^{10} \approx 60\%$ while the false positive rate will be $0.52^{10} \approx 0.14\%$. 

The training process of single classifier is aiming to minimize errors, not to achieve high detection rates at expense of large false positive rates. A straightforward trade off is to adjust the threshold of the decision rule produced by AdaBoost. In general, lower threshold yields classifier with more false positives and higher detection rate; higher threshold yields classifier with fewer false positives and a lower detection rate. 

In the implementation of the algorithm, there are some parameters need to be tuned carefully.  Our final algorithm will be able to combine the following factors together and optimize the running time while achieving the detection rate and false positive rate goal, we have more parameters to handle than Viola and Jones \cite{Viola01robustreal-time}:
\begin{itemize}
	\item Number of stages
	\item Number of features of each stage
	\item Threshold for each stage
	\item The cutoff threshold for \emph{high value}
	\item The cutoff threshold for \emph{significant} price separation
\end{itemize}
 
In the implementation, we manually set up the bounds for maximal accepted fpr (false positive rate) and minimum accepted detection rate or tpr (true positive rate). The reason we involve a manual process is that to find an optimum combination of the above numbers is extremely hard, it does not worth putting the computation resource. Each layer of cascade is trained by AdaBoost with increasing number of features until the detection and false positive rates are met for that stage. Then both rates are confirmed by testing on a validation data set. We also evaluate the overall false positive rate after finishing training and testing one level, if the bound is not yet met, then another layer is added to the cascade. We put the false detections on the true negative set into the training set of subsequent classifiers. The two cutoff thresholds also have a direct impact on seller revenue, we run a finite grid search for the optimal values, the results are shown in Section \ref{sec: simu results}. 

\begin{footnotesize}
\begin{table}
\caption{Pseudo code for cascading algorithm.}
\label{cascade code}
 \begin{tabular}{ll}  
\hline
1& $F_i = $ fpr of combined classifier after $i$ rounds \\
2& $D_i = $ tpr of combined classifier after $i$ rounds \\
3& $f = $ maximal acceptable fpr, $d = $ minimal acceptable tpr \\
4& $F_{TGT} = $ target fpr \\
5& $P = $ positive examples, $N = $ negative examples \\ \hline
6& Initialize $F_0  =  1, D_0=  1, i=0 $ \\
7& while $F_i > F_{TGT}:$  \\
8&	$\quad$ $i++, C_i=0, F_i=F_{i-1}$ \\
9&	$\quad$ while $F_i > f*F_{i-1}$: \\
10&	$\quad \quad$ $C_i++$ \\
11&	$\quad \quad$ train a classifier with $C_i$ features using $P$, $N$ \\
12&	$\quad \quad$ calculate $F_i$ and $D_i$ \\
13&	$\quad \quad$ decrease threshold for $i$th classifier s.t $D_i\geq d*D_{i-1}$ \\
14&	$\quad$ empty $N$ \\
15&	$\quad$ if $F_i > F_{TGT}$: \\
16&  $\quad \quad$ apply classifier on true negative samples and put false prediction into $N$ \\
\hline
\end{tabular}
\end{table}
\end{footnotesize} 

Note that in line $13$ of Table \ref{cascade code}, updating threshold also impacts $F_j$, so there is a risk to have a endless loop. In practice, this is not a problem to us since we have a realistic $F_{TGT}$ with our cascading classifiers.

\section{Simulation Results and Conclusions}  \label{sec: simu results}

One simple and fast way of implementing this model for testing is to enforce the predicted floor prices in selling rule for a portion of the traffic. For a online A$|$B test, we update the selling rule in the keys that have the following six dimensions: hour, gob  (group of buyers), siteid, adsize, adpos, state. In our system, more than $80\%$ of the traffic condensed on about $1200$ keys. We noticed that on those keys, high-value inventories are highly condensed, and these $1200$ keys consists of about $60\%$ of the total high-value traffic. The revenue lift is around $3.5\%$ using our trained classifiers in the cascade model. When underbid happens, we actually route part of the blocked traffic to other platforms for reselling. The revenue from blocked traffic is provided by the third party with delay, so we do not include this part of revenue in our simulation results.

\begin{footnotesize}
\begin{table}
\caption{Simulation Results on 10\% Random Data Sample. }
\label{table: simu res}
\begin{tabular}{|l|l|l|l|}  
\hline
  \textbf{Effected Auctions} & High-value Auctions & Low-value Auctions & Total \\ \hline
  Current Revenue & \$30,626  & \$85,753   & $\$116,379$ \\  \hline
  \emph{New Revenue} & $\emph{\$40,316}$  & $\emph{\$85,647}$  & $\mathbf{\$125,963}$  \\  \hline
  \cline{1-4}
  \textbf{Un-effected Auctions} & -  & -  &  $\$160, 761$   \\  \hline \cline{3-4}
  \multicolumn{2}{l|}{} & Global Revenue Lift & \textbf{3.5\%} \\ \cline{3-4}    
\end{tabular}
\end{table}
\end{footnotesize}

  
In Table \ref{table: simu res}, we show one simulation results on a random 10\% sample and are able to achieve around \$1000 lift on the effected auctions (the revenue on the un-effected auctions left unchanged), which converts to a 3.5\% global lift. The portion of high-value inventories in the marketplace is relative small. For instance, auctions with top bid above $\$10$ is about $5\%$, but it contributes 40\% total revenue. Although the revenue lift is around $3.5\%$ for the marketplace, it is pretty stable over a long period with fluctuations around $0.3\%$. Given the size of our exchange revenue on display ads, it translates to an extra two dozens of million dollars in a yearly basis. Currently, we are slowly push the model to the production with small traffic, as of now we have used only three level of cascading to reach this, and we implement our model in an extremely conservative way. Usually the higher false positive results in higher resell rate, in order to maintain certain marketshare level we put a \emph{firm} constrain on the fpr. As a consequence, we sacrifice the detection rate as much as possible to meet this goal even though we can further grow our sell revenue by abandoning this constrain. 

Future concerns about this system may include studying of the long term behavior of bidders, it is possible that some bidders will shade their bids after noticing that they are paying a higher prices for this high-value inventories. However, after bid shadings, bidders may lose some auctions and potentially boost the shaded bids to a higher level. It is hard to predict the consequence of this mechanism, but it will definitely put some effects on the total revenue we can draw from this model.

Another note we would like to add is that there is still large potential for the cascading model to achieve better revenue performance. Due to engineering limitations, we haven't make fully use of Yahoo's user profile data in this model. Some important user features that we used in ads targeting are absent, such as the user segment, content taxonomy, beacon events, etc. By adding those feature, we may build more accurate classifiers. 

\section{Acknowledgements}
We would like to acknowledge the help from former Yahoo! Inc. employee Yu Wang for a lot of useful discussions and the implementation of the test environment.

\bibliographystyle{plain}
\bibliography{optRes} 

\end{document}